\documentclass[aps]{revtex4}

\usepackage{epsfig}

\def\e{\mathop{\rm \mbox{{\Large e}}}\nolimits}
\newcommand{\bn}[1]{\mbox{\boldmath $#1$}}
\newcommand{\be}{\begin{equation}}
\newcommand{\ee}{\end{equation}}
\newcommand{\bc}{\begin{center}}
\newcommand{\ec}{\end{center}}
\newcommand{\bea}{\begin{eqnarray}}
\newcommand{\eea}{\end{eqnarray}}
\newcommand{\ba}{\begin{array}}
\newcommand{\ea}{\end{array}}

\begin{document}

\title{Symmetries and dynamics in an AC-driven self-assembled quantum dot lens}
\author{Arezky H. Rodr\'{\i}guez$^{1}$, Lilia Meza-Montes$^{2}$, Carlos Trallero-Giner$^{3}$
and Sergio E. Ulloa$^{4}$}

\affiliation{$^{1}$ Universidad Aut\'onoma de la Ciudad de M\'exico (UACM), Ermita
Iztapalapa 4163, Col. Lomas de Zaragoza, C.P. 09620, Iztapalapa, M\'exico D.F., M\'exico. \\
$^{2}$ Instituto de F\'{\i}sica, Universidad Aut\'onoma de Puebla, Apdo. Postal J-48, Puebla, Pue.
72570, Mexico \\
$^{3}$ Departamento de F\'{\i}sica Te\'orica, Universidad de La Habana, Vedado 10400, Cuba \\
$^{4}$ Department of Physics and Astronomy, Condensed Matter and Surface Science Program, Ohio
University, Athens, Ohio 45701-2970, USA}

\date{\today}

\begin{abstract}
Theoretical results for a single electron in multi-level system given by a lens-shape
self-assembled quantum dot in the presence of an intense harmonic electric field are presented. A
non-perturbative Floquet approach is used to study the dynamical localization of the particle when
going beyond the two-level approach by introducing the full spectral level structure. It is
discussed the role of the different quasi-energy sidebands as the parameters of the system change.
It is found that the contribution of different drive harmonics is controlled by fine tuning of
field intensity. It is also shown that avoided crossings in the quasi-energy spectrum are
correlated with the spectral force of the sidebands and dynamical state localization.
\end{abstract}



\maketitle

\section{Introduction}

There is intense activity on the experimental and theoretical understanding of the dynamical
evolution of quantum systems exposed to strong time-dependent external fields
\cite{shirley,grifoni1998}. The topic has acquired further relevance in connection with the
practical operation of devices subjected to oscillating electrical and magnetic fields at the
nanoscale. Examples include the shift of resonances in heterostructures as ac-fields are applied
(the ac-Stark effect) \cite{trallero2004}, the behavior of electronic bands in spatially periodic
systems \cite{milfeld,Schulz2002,Martinez2002,platero2001}, and the production of currents in an
ac-driven quantum dot \cite{platero2004-1}. One important effect in these systems is the strong
dynamical suppression of tunneling at suitable values of applied ac-field. The coherent destruction
of tunneling that appears, known as dynamical localization in the literature, has been well studied
in two-level systems as coming from the destructive interference introduced by the drive
\cite{grossmann,grifoni1998}, whenever there is a crossing of quasi-energy levels in the spectrum.
Dynamical localization has been proposed as a tool to control the spatial location of a particle in
a two-well potential \cite{grifoni1998}, and to selectively control the tunneling in a
multiple-well system \cite{Ulloa2004}.

Typical growth conditions of semiconducting quantum dots result in dots with lens geometry, and an
analysis of this spatial symmetry on the electronic structure is of interest
\cite{Yoffe2001,jpacm}. In this work we explore the problem of periodic driving force and dynamical
localization in a realistic level structure that describes self-assembled quantum dots in
semiconductors. This requires that we extend the Floquet formalism to self-assembled quantum dots
with lens shape. We also establish the importance of incorporating the multi-level structure of a
real system and identify in this complex level structure the conditions for dynamical localization.

We find the realistic lens shape to be crucial in the description of the dynamics, as the spatial
non-separability of the state plays a relevant role. Consideration of the multilevel structure
present in typical quantum dots is a vital requirement for the correct description of the dynamical
response of carriers. It is essential to go well beyond the consideration of only two active levels
to fully describe the Floquet quasi-energy and the time evolution of electrons in realistic quantum
dots. Even for weak driving forces or frequencies, the description of the dynamics requires the
inclusion of many different states in order to achieve a fully converged description of the time
evolution. The quantum lens geometry makes for a complex and rich theoretical description of the
problem \cite{jap}. Interestingly, we show that the real dot system allows the generation of higher
harmonics of the driving frequency, with intensity that is fully dependent on the amplitude of the
drive, which could be used for its generation. The phenomenon of dynamical localization is shown to
remain for suitable values of driving field, with strongly diminished localization at high
intensity fields.

\section{Formalism and Result.}

We consider a typical self-assembled quantum dot (SAQD) with lens symmetry of circular cross
section of radius $a$ and maximum height $b$ which is harmonically driven by an electric field
along the axial symmetry $z$ of the lens with intensity $F$ and frequency $\omega$, $\bn{F} = F
\sin\omega t \; \bn{\hat{z}}$. Assuming that the electron is described by an isotropic band with
effective mass $m^*$, the dynamics of the system is governed by the time-dependent Schr\"{o}dinger
equation
\begin{equation}
\label{eq1}
\widehat{L} \, \Psi(\bn{r},t) = \left( -\frac{\hbar^2}{2m^{\ast}} \nabla^{2} - e F \bn{\hat{z}}
\cdot \bn{r} \sin\omega t \,- i \, \hbar \, \frac{\partial}{\partial t} \right) \,\Psi(\bn{r},t) =
0,
\end{equation}
where, in spherical coordinates, we have $\bn{\hat{z} \cdot \bn{r}} = r \cos\theta$. The space of
functions where the operatior $\widehat{L}$ of Eq. (\ref{eq1}) is defined, corresponds to those
spatio-temporal functions which are bounded functions defined in the real space $\mathcal{R}_{3}$
of the lens domain and are also periodic functions on time with period $\tau =2\pi /\omega$. The
solution of Eq. (\ref{eq1}) can be obtained following the standard Floquet theory
\cite{grifoni1998} where $\Psi(\bn{r},t)$ is written as
\begin{equation}
\label{eq2}
\psi (\bn{r},t) = \e^{\textstyle{-i \,\varepsilon \, t / \hbar}} \; \varphi (\bn{r},t),
\end{equation}
which allows to rewrite Eq. (\ref{eq1}) as an eigenvalue problem with the same operator
$\widehat{L}$ for a real-valued eigenenergy $\varepsilon$ (called quasi-energy) and eigenfunction
$\varphi (\bn{r},t)$ which fulfills the periodic condition $\varphi(\bn{r},t) = \varphi(\bn{r},t +
\tau)$. The periodic time part of the function is expanded as
\begin{equation}
\label{eq4}
\varphi (\bn{r},t) = \sum_{n = - \infty}^{\infty} \frac{ \e^{\textstyle{i\,n\,\omega \,t}}
}{i^{n}\sqrt{2\pi /\omega }} \, u_{n}(\bn{r}).
\end{equation}
Furthermore, the function $\varphi_n = \e^{\textstyle{i n \omega t}} \varphi$ is also a solution
with quasi-energy  $\varepsilon_n = \varepsilon + n \hbar\omega$ \cite{shirley,grifoni1998}. These
additional solutions, a consequence of the time periodicity, have been called replicas or
sidebands. By subtracting a suitable integer multiple of $\hbar\omega$, the quasi-energy
$\varepsilon$ is mapped onto the first Brillouin zone (FBZ) which, in units of $\hbar\omega$,
will be scaled to [0,1] for convenience. As the ac-field is selected along the $z$-axis, the
$z$-component of angular momentum with quantum number $m$ is preserved. Then, the spatial component
$u_{n}(\bn{r})$ in each subspace with a given value of $m$ is expanded as a linear combination of
the complete set of functions $\left\{ \Phi^{(b/a)}_{N,m} \right\}$ for the tridimensional lens
domain, as reported in Ref. \cite{jpacm}. Finally it is obtained an infinite eigenvalue problem for
the quasi-energy $\varepsilon$ and weight coefficients $C_{n,N,m}$. The complexity of the problem
depends on the values of the field intensity $F$ and frequency $\omega$. The solution is sought in
terms of a truncated basis set ($N = 1, ..., N_{max}$) which is made as large as needed in order to
reach the desired convergence. Along this paper it is used a lens domain with ratio $b/a=0.71$, 25
energy levels in the expansion of $u_{n}(\bn{r})$, and for each energy level, the index $n$ for the
replicas in Eq. (\ref{eq4}) was taken as $n=0, \pm 1, \pm 2, ..., \pm 20$.

Starting from an initial state at $t=0$, the carrier particle is induced to explore the complete
spectrum of the system when the ac-field is connected. How this occurs as a function of time can be
analyzed by following the evolution of an initial electron state $f_o(\bn{r}) = f(\bn{r},t=0)$,
which is written as a linear combination of Floquet states
\begin{equation}
\label{eq16}
f(\bn{r},t) = \sum_{\{P\} \in FBZ} A_{P} \; \Psi _{P}(\bn{r},t),
\end{equation}
with coefficients $A_{P}$ fixed by the initial conditions. The summation is taken on the First
Brillouin Zone (FBZ), including all the replicas \cite{shirley}. At $F=0$ the label $P$ indicates
the index of quasi-energy $\varepsilon_{N,n}$ with $n$th replica and level $N$. For simplicity we
consider that the initial state has well-defined $z$-component of the angular momentum in such a
way that only states with the same $m$ are considered. Then, Eq.\ (\ref{eq16}) can be cast in the
following way
\begin{equation}
\label{eq17}
f_{m}(\bn{r},t) = \sum_{N} \Delta _{N}(t) \Phi_{N,m}^{(b/a)}(\bn{r}),
\end{equation}
where
\begin{eqnarray}
\label{eq18}
\Delta _{N}(t) = \sum_{\{P\} \in FBZ} \sum_{n=-\infty}^{\infty} A_{P} \; C_{n,N,m}(P) \;
\frac{\e^{\textstyle{i (n\omega t-\pi /2)}}}{\sqrt{2\pi/\omega}} \, \e^{\textstyle{-i \,
\varepsilon(P) \, t /\hbar}}
\end{eqnarray}
The function $\Delta _{N}(t)$ contains all the dynamical information due to the presence of the ac
electric field. Thus, $P_{N}(t)=\left\vert \Delta _{N}(t)\right\vert ^{2}$ gives, at a given time
$t$, the probability of finding the system in the state $\Phi _{N,m}^{(b/a)}$. The typical
definition of dynamical localization considers that the carrier does not evolve away from its
initial spatial state or configuration, as the effective tunneling amplitude from site to site is
suppressed \cite{grifoni1998}. Here, we monitor dynamical localization by the corresponding quantum
probability of finding the particle in its initial state as a function of time. In what follows we
assume that the initial state $f_o(\bn{r})$ is the zero-field ground state
$\Phi_{N,m}^{(b/a)}(\bn{r})$ with $N=1$ and $m=0$.

\begin{figure}[t] 
\vspace*{3cm}
\centerline{\psfig{file=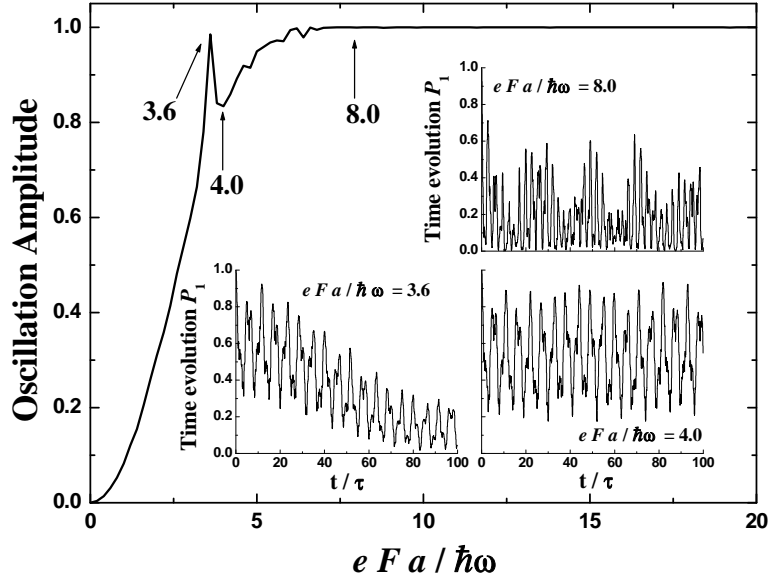,width=0.6\textwidth}}
\vspace*{-2.5cm}
\caption{Oscillation Amplitude as a function of the dimensionless field intensity $e F a / \hbar\omega$
for $b/a=0.71$ and $\Omega = 100$. Insets show the probability $P_1$ of finding the particle in the
initial state as a function of time for different values of $e F a / \hbar\omega$.}
\label{fig1}
\end{figure}

\begin{figure}[t] 
\vspace*{3cm}
\centerline{\psfig{file=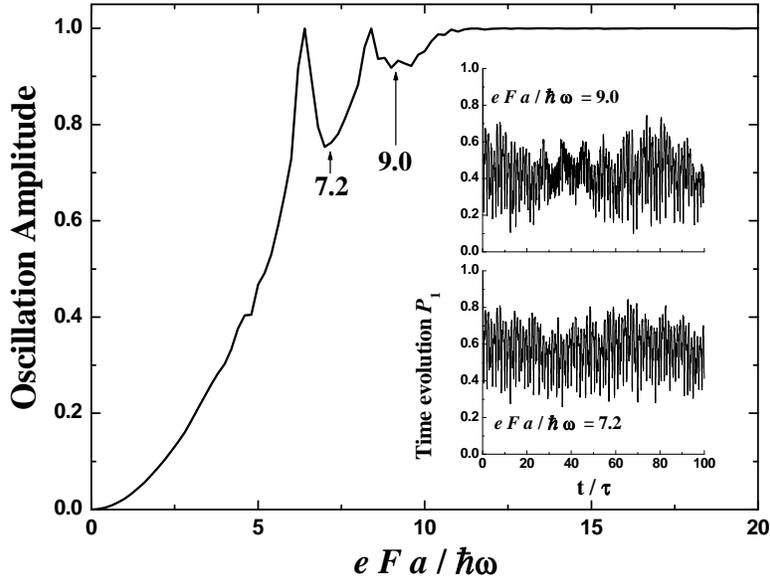,width=0.6\textwidth}}
\vspace*{-2.5cm}
\caption{Same as in Fig. \ref{fig1} for $\Omega=300$.}
\label{fig2}
\end{figure}

In order to quantify the degree of localization, we study the Oscillation Amplitude (OA) as a
function of the field intensity, where the OA is defined as $OA = P_{1}(t=0) - P_{min}$, where
$P_{min}$ is the minimum value that $P_{1}(t)$ takes over a long interval of time (100 time units
in our case) \cite{creffield-platero}. In Fig. \ref{fig1} and \ref{fig2} the OA is plotted as a
function of the dimensionless parameter $e F a / \hbar\omega$ where $\Omega =\hbar \omega / E_{o}$
is a dimensionless frequency with $E_{o} = \hbar^2 / (2 m^* a^2)$ as a unit of energy. At zero
field the OA is naturally always zero and increases with the field strength, since the latter
induces mixing of the states and it forces the system to explore larger regions of the eigenvalue
spectrum of the system away from the initial state. The OA reaches the maximum value ($OA = 1$) at
certain value of the reduced field intensity and for $e F a/\hbar\omega > 11$ it is near unity,
indicating that the probability of the system remaining in the initial state is zero. However, the
OA strongly decreases at some particular field intensities (for example at $e F a/\hbar\omega = 4$
in Fig. \ref{fig1} and at $e F a/\hbar\omega = 7.2$ and 9.0 in Fig. \ref{fig2}), meaning that
$P_{1}(t)$ never goes to zero at those values of the field intensity, and quasi-localization of the
system can be identified. At the same time and to study explicitly the behavior of $P_1$ on time,
different panels inside Fig. \ref{fig1} and \ref{fig2} show the time evolution $P_1(t)$ during one
hundred time units $\tau =\hbar /E_{0}$ at different values of the dimensionless field intensity.
According to Eq. (\ref{eq18}), the wave function is a multi-periodic function and presents strong
oscillating behavior as seen in the inset of the figures. Notice, moreover, that at small value of
reduced frequency ($\Omega = 100$), the mixture of the spectrum is strong for $e F a / \hbar\omega
= 3.6$ in Fig. \ref{fig1} and for $e F a / \hbar\omega = 4$ a condition for quasi-localization is
reached. Instead, for higher values of the frequency ($\Omega = 300$), a lower slope for the
increase of the OA is obtained as seen in Fig. \ref{fig2}. It is also reported two values for
quasi-localization at $e F a / \hbar\omega = 7.2$ and 9.0. Notice also that the stronger the
intensity the lower the values of the probability, a consequence of the strong level mixing as
discussed previously.

\begin{figure}[t] 
\vspace*{-1cm}
\centerline{\psfig{file=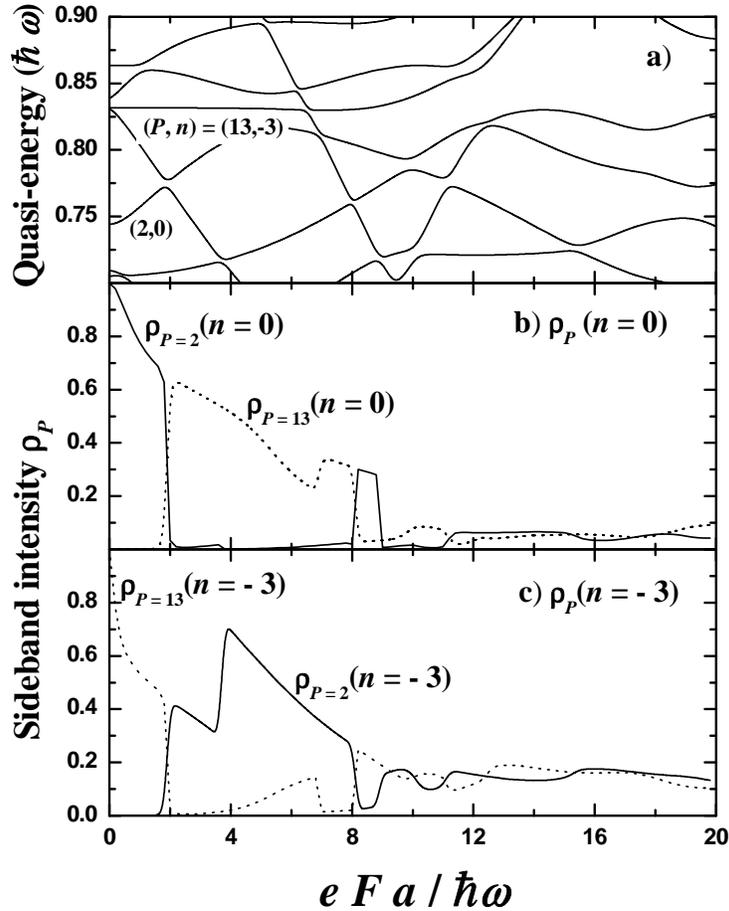,width=0.6\textwidth}}
\vspace*{-1cm}
\caption{a) Small window of the FBZ for the quasi-energies as a function of the reduced field intensity.
b) Intensity sidebands $\rho_{P}(n)$ for $P=2$ and $P=13$ with $n=0$. c) Idem as b) but now with
$n=-3$. Calculations are done for $b/a=0.71$ and $\Omega = 100$.}
\label{fig3}
\end{figure}

On the other hand, the interaction among different lens levels in the quasi-energies as function of
the field amplitude can be analyzed in more detail using the sideband intensity or spectral force
$\rho_P$, which amounts to the weight of the $n$-th sidebands on the $\psi _{P}(r,t)$ Floquet
state. Thus, according to the expansion of $u_n(\bn{r})$, the spectral force is given by
\begin{equation}
\label{eq14}
\rho_P(n)=\sum_{N}|C_{n,N,m}(P)|^{2}.
\end{equation}
Notice that at $F=0$, $\rho_P(n) \equiv \rho_N (n)$. On the other hand, for $F>0$, the spectral
force of a given $P$ indicates the weight of the different lens states in the full Floquet
expansion (\ref{eq4}). Figure \ref{fig3} a) shows a small window of the quasi-energy spectrum,
where some anticrossings are present at $\Omega = 100$. The system parameters are the same as in
Fig. \ref{fig1}. In Figs. \ref{fig3} b) and \ref{fig3} c) the intensity sidebands $\rho_P (n)$ for
the quasi-energies $P=2$ and $P=13$ have been plotted, with $n=0$ in Figs. \ref{fig3} b) and $n=-3$
in Figs. \ref{fig3} c) respectively. We observe in the figures the effects in the spectral force
$\rho_p(n)$ caused by the the anticrossing between quasi-energies $P=2$ and $P=13$ around $e F
a/\hbar \omega \simeq 1.8$. For values up to $eFa/\hbar \omega \simeq 1.8$, $\rho_{P=2}(n=0)$ and
$\rho_{P=13}(n=-3)$ show a large intensity as expected due to its corresponding $F=0$ limit, while
$\rho_{P=13}(n=0)$ and $\rho_{P=2}(n=-3)$ are barely noticeable near $F\simeq 0$. However, at the
corresponding anticrossing, a strong mixing of the states takes place, and $\rho_{P=13}(n=0)$ and
$\rho_{P=2}(n=-3)$ increase rapidly while $\rho_{P=2}(n=0)$ and $\rho_{P=13}(n=-3)$ strongly
decrease for $eFa/\hbar \omega > 1.8$, i.e., their corresponding strengths are inverted after the
anticrossing. Thus, the spectral weights for $n=0$ and $n=-3$ in this case, and in general for all
the spectral contributions, are exchanged between the Floquet states at the anticrossing. Most
importantly, as the field increases, so does the interlevel mixture and the weights of the various
replicas become nearly identical, so that the amplitude of their contribution are similar. We
emphasize that this nearly homogeneous distribution of the spectral force over many different
replicas results in substantially different time evolution of the driven system. We then anticipate
that time averages of physical observable could be substantially different than when only the
lowest two lens levels are considered \cite{Ulloa2002}.

\section{Conclusions}

We have analyzed the time evolution of an electron in self-assembled quantum dots with lens shape
in the presence of intense radiation along to the rotational axis of the lens. We exploit this
axial symmetry of the lens domain to solve this complex time-dependent problem. This realistic
driven single-electron system has been studied over a wide range of field amplitudes. We have
demonstrated that consideration of the typical two-level approximation yields an incomplete
description even at moderate fields and frequencies. We have also calculated the complex
quasi-energy spectra that result in this problem, and analyzed the anticrossings that appear in
terms of the interaction among zero-field states and their replicas. We find that these
anticrossings are associated with strong shifts in the spectral weights for the Floquet states
between two quasi-energies, and that for larger field intensities the spectral weights are
distributed homogeneously among a wide range of sidebands. This strong dependence indicates that
the appearance of different drive harmonics in the response of the system could be easily
controlled by the field strength. It is interesting to consider the possibility of utilizing such
lens shape quantum dots in strong ac-fields as a source of different harmonics.

In order to study whether dynamical localization prevails under these much more complicated
conditions of multi-level dynamics, we have studied the time evolution of the system prepared
initially in the zero-field ground state. We find that at field intensities for which a
quasi-energy anticrossing appears in the spectrum, some degree of state localization is observed;
in those circumstances, the probability of finding the system in the initial state never goes
completely to zero, but it reaches a minimal value. This incomplete dynamical localization has been
analyzed quantitatively by means of the oscillation amplitude for different system parameters.
Similar to the case of two-level systems, we find that the dynamical localization, already
precarious at high frequency, disappears for lower frequency values. The analysis and results we
present are important for the interpretation of experimental data, and suggest further theoretical
work to assess the relevance of multi-level structures in realistic systems.


\end{document}